Blue hypertext is a perfect design decision: No perceptual disadvantage in reading and successful highlighting of relevant information.

Benjamin Gagl

Department of Psychology, Goethe University Frankfurt & Center for Individual Development and Adaptive Education of Children at Risk (IDeA), Frankfurt am Main

Address for correspondence:

Benjamin Gagl

Department of Psychology

Goethe University Frankfurt

Theodor-W.-Adorno-Platz 6

60323 Frankfurt/Main, Germany

Tel.: 0049 69 798 35336

E-mail: gagl@psych.uni-frankfurt.de

Word count: 2429 (excluding Abstract and References)

Abstract

Highlighted text in the Internet (i.e. Hypertext) is predominantly blue and underlined. The percept of these hypertext characteristics were heavily questioned by applied research and empirical tests resulted in inconclusive results. The ability to identify blue text in foveal and parafoveal vision was identified as potentially constrained by the low number of foveally centered blue light sensitive retinal cells. The present study investigates if foveal and parafoveal perceptibility of hypertext is reduced during reading. A silent-sentence reading study with simultaneous eye movement recordings and the invisible boundary paradigm, which allows the investigation of foveal and parafoveal perceptibility, separately, was realized. Target words in sentences were presented in either black or blue and either underlined or normal. No effect of color and underlining, but a preview benefit could be detected for first pass reading measures (comparing fixation times after degraded vs. un-degraded parafoveal previews). Fixation time measures that included re-reading (i.e., total viewing times) showed, in addition to a preview effect, a reduced fixation time for not highlighted (black not underlined) in contrast to highlighted target words (either blue or underlined or both). Thus, the present pattern reflects no detectable perceptual disadvantage of hyperlink stimuli but increased attraction of attention resources, after first pass reading, through highlighting. Blue or underlined text allows readers to easily perceive hypertext and at the same time readers re-visited hypertext longer as a consequence of highlighting. On the basis of the present evidence blue hypertext can be safely recommended to web designers for future use.

Keywords: Reading, Hypertext, Eye movements, Invisible boundary paradigm, blue color, underlining.

Introduction

The Internet plays an incredibly important role in our daily life. One of the first but also most critical advantages of the Internet is the use of hypertext. Hypertext allows the web designer to efficiently link important snips of text to additional information. Thus, hypertext works by replacing referencing in printed text and eliminating effortful literature searches. The most common implementation of hypertext, embedded as blue underlined text, was prominently criticized (Nielson, 1999). It was argued that choosing blue as text color is a poor choice as only about 2% of retinal cells are sensitive to blue. As a consequence, blue hypertext might reduce reading speed due to hampered foveal processing. This would be unfortunate since it would limit the general increase of effectiveness of text processing introduced by hypertext. In addition, color sensitive retinal cells are most prominent in the fovea of the eye and their number decreases massively towards para- and extra-foveal regions of the retina. This massive reduction of color sensitive cells towards the para- and extra-foveal regions might also decrease the possibility to extract relevant parafoveal information from colored text in reading. In general, parafoveal preprocessing typically increases reading speed drastically (for a review see Schotter, Angele, & Rayner, 2012). Therefore a decrease in reading speed resulting from both reduced parafoveal preprocessing and foveal perception would be drastic when considering how often hypertext is read daily. Such a decrease in reading efficiency would indicate that the use of blue underlined hypertext cannot be recommended.

Recently, Fitzsimmons, Weal, and Drieghe (2013) investigated the influence of colored words on eye movement measures and found a reduced skipping probability (i.e., the probability of not fixating a word) of blue words. Their paradigm allows access of the combined foveal and parafoveal processing during silent reading of sentences. They found a reduced reading speed, in contrast to black text, for words written in gray but not for words written in other colors (e.g., blue). This finding indicates that contrast (black vs. gray) but not color (e.g., black vs. blue) hampers reading speed. For words presented in saturated colors (e.g., blue) they found a reduced skipping probability in contrast to black-presented words. The latter finding can be interpreted in two ways: Either bottom-up perceptual processes are hampered due to a reduced parafoveal perceptibility of blue words, increasing the fixation probability. Or top-down processes increase the fixation probability reflecting the learned association of hypertext to informative content attracting additional attentional resources to highlighted words.

To differentiate between these interpretations the present study realized an invisible boundary paradigm (Rayner, 1975; for a revised version see Gagl, et al., 2014). This paradigm allows researchers to estimate the parafoveal preview benefit by contrasting fixation times after perfect previews (no manipulation) in contrast to degraded previews (limiting preview benefits). The task of the participant is reading sentences silently as if they were reading a book or newspaper (i.e. as natural as possible). An invisible boundary is placed before a target word (see Figure 1a). When the invisible boundary is crossed by a saccade the change from a degraded to an un-degraded target word presentation is realized during the eye movement. The increase of reading speed after the parafoveal presentation of a normal word compared to the condition with a degraded word is interpreted as the parafoveal preview benefit. The boundary paradigm cannot be optimally implemented in case the skipping rate is expected to vary drastically between conditions, as the estimation of the preview benefit relies on the fixation times on the target word. To realize high target word fixation rates, the predictability out of the sentence context was held low for the target words, which decreases skipping probabilities (Fitzsimmons, & Drieghe, 2013; Hawelka, Schuster, Gagl, & Hutzler, 2015). Therefore, low skipping rates, at the best-case floor effects, are expected to reduce the probability of finding differential effects in this measure. The effects of the present manipulations are expected in the fixation time measures of first fixation duration (i.e. the duration of the initial fixation), gaze duration (i.e., the summated fixation duration of all fixations during the first encounter) and total viewing time (i.e., the sum of the gaze durations plus the fixation durations after regressive saccades to the target words; re-reading). To investigate the highlighting hypothesis (i.e. more top-down attention is allocated to highlighted text), in addition to the preview manipulation and the color manipulation, highlighting was manipulated separately by underlining. The resulting design included the factors color (blue vs. black), underlining (underlined vs. not underlined) and degradation (degraded preview vs. un-degraded preview; see Fig. 1).

In case parafoveal bottom-up processing of blue stimuli is limited, a reduced parafoveal preview benefit in contrast to black words is expected. Limited foveal bottom-up processing of blue text would result in higher fixation times of blue vs. black target words. This should be the case irrespective of underlining or parafoveal preview. Both parafoveal and foveal findings would indicate a hampered bottom-up processing of blue hypertext. Alternatively, if top-down processes that originate from highlighting influence the reading behavior, than the un-highlighted condition (i.e., black not-underlined targets) should receive less attention. In contrast, the highlighted words, blue not-underlined, blue underlined and black underlined targets, should receive additional attention reflected in longer fixation duration measures.

## Methods

### Participants

Forty native German–speaking students (24 female; mean age: 23:2 years:month; standard deviation: 2:0) with normal reading speed measured by the unpublished adult version of the Salzburger-Lese-Screening (SLS; Auer, Gruber, Mayringer & Wimmer, 2004; for the current state of the adult version see Gagl, Hawelka, & Hutzler, 2014) and normal or corrected-to-normal vision participated. One additional participant was excluded due to very slow reading (Percentile < 16).

Apparatus

Movements of the right eye were recorded with a sampling rate of 2,000 Hz (EyeLink CL eye-tracker, SR-Research, Canada). Participants were seated about 52 cm in front of a CRT monitor (150-Hz refresh rate; screen resolution of 1024x768 pixels) and a forehead and chin rest stabilized their heads. The display change latency of the experimental setup was below 15 ms (for details see Richlan et al., 2013).

Material

The manipulation of color and highlighting was realized with five letter target words embedded in sentences, which were matched on the most important word characteristics (e.g., orthographic similarity: OLD20, Yarkoni et al., 2008; word frequency: SUBTLEX database, Brysbeard et al., 2011; and predictability from sentence context, e.g., Kliegl, Grabner, Rolfs & Engbert, 2004). Furthermore eight different versions of the sentences (N = 320) allowed the presentation of each sentence in one of the eight conditions (n = 40; Fig. 1a). An equal number of participants were assigned to each version (n = 5). The parafoveal preview manipulation was realized by randomly replacing 45% of the black or blue pixels of the presented letters (for details see Gagl, et al., 2014). This procedure distorted the parafoveal percept of the target words without inhibiting lexical processing. The sentences were presented in a mono-spaced font (single character width: 0.3° of visual angle) and target words were never at the first, second, or final position of the sentences.

Procedure

A 3-point calibration of the eye tracker preceded the experiment. Fixating between two vertical lines in the left margin of the monitor triggered sentence presentation in such a way that the participants' fixation was at the center of the sentence's first word. The students read silently for comprehension. After, on average, a quarter of the sentences, the experimenter orally presented comprehension questions, which the participants almost always answered correctly (M = 96%).

All words after the target word were visually degraded to minimize potential influences of these words (i.e., particularly of n+2, with n+1 being the target word; see Kliegl, Risse & Laubrock, 2007). After crossing the invisible boundary at the end of the pre-target word, the target word and the remainder of the sentence were presented un-degraded (see Fig. 1a). Fixating an 'x' in the lower right corner of the screen terminated the trial. Ten practice trials preceded the experiment. Recalibration was conducted after the practice trials, after a break halfway through the experiment, and when the fixation control at the start of a trial failed.

Data Treatment and Analyses

Skipping probabilities, first fixation durations, gaze durations, and total viewing time are reported. First fixation durations, gaze durations and total viewing times shorter than 80 ms were removed from the data (for each measure < 1% of the data). Data analysis was administered with linear mixed effect models (LMMs) for the *log*-transformed fixation timing measures and generalized linear mixed effect models (GLMMs) for the skipping probability (this analysis is best suited to estimate binary data: skipped vs. fixated) with the lme4-package (Bates, et al., 2015) in R. G/LMMs are suited for analyzing unbalanced data (e.g., due to skipping of target words). Color, underlining, degradation and all interactions were included in the models as fixed effects. Random effects were estimated for the intercepts of both participants and items. In addition, the random slopes for the fixed factors were added to the model until an additional parameter did not allow the model to converge. In case adding another level to the random effect structure resulted in a not converging model, one of the other two factors was introduced into the model and the model was refitted. If two models with the same number of random slope estimates converged, an ANOVA was used to compare the model fits and allowed to decide which model estimated the data better. This procedure resulted in the additional estimation of the random slope of color on the random effect of participant for the skipping probability. For the first fixation duration, the random slopes of degradation and color were estimated for the random effect of participant. For the gaze duration, the random slopes of underlining, color and degradation on the random effect of participant and the random slope of degradation on the random effect of item were estimated. For the total viewing time the random slopes of underlining, color and degradation on the random effect of participant were estimated. With this procedure, the most conservative converging models were selected.

Results and Discussion

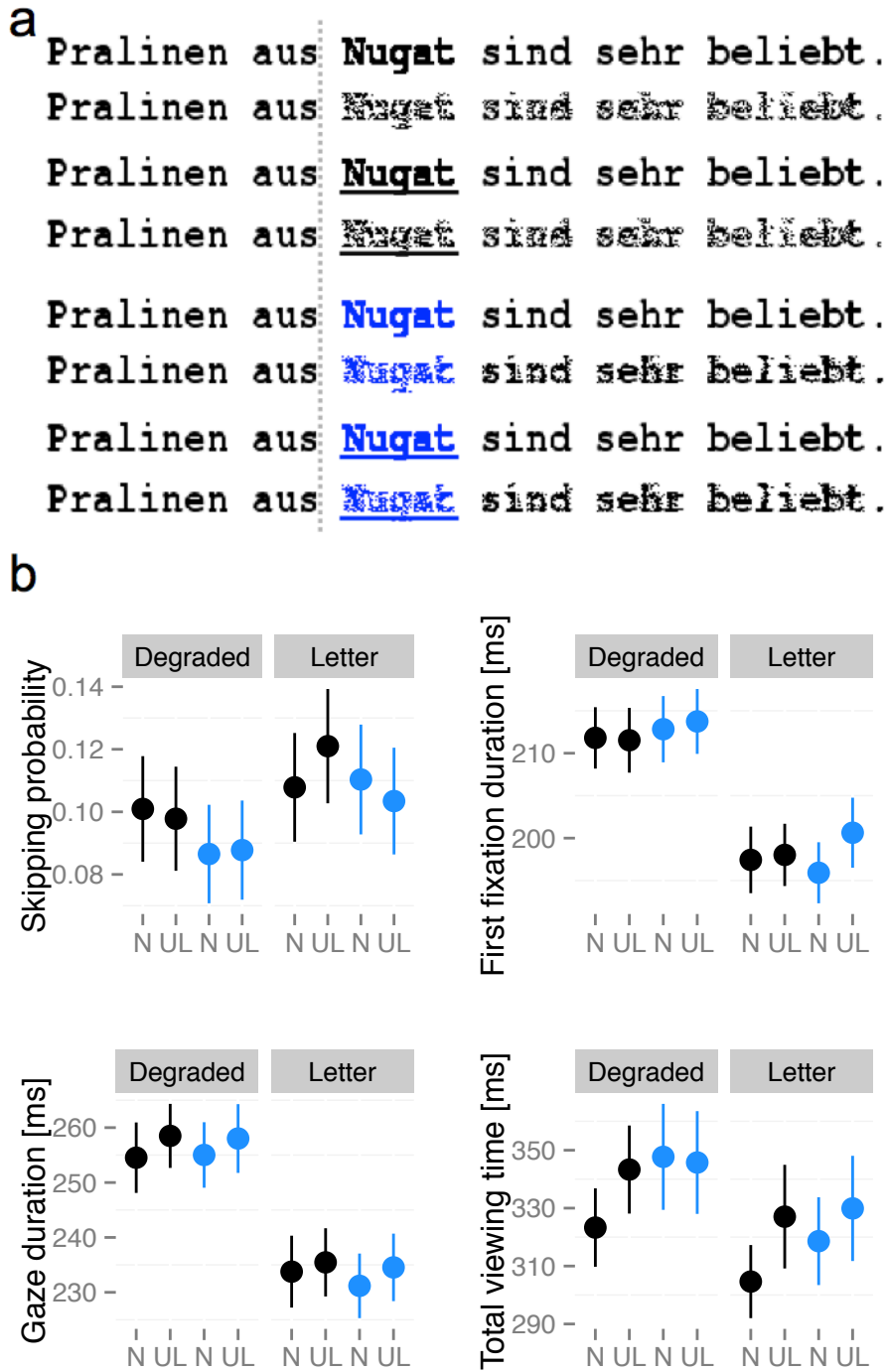

Figure 1. Stimulus presentation and eye movement data. (a) Example sentence for all eight conditions (black not-underlined, black underlined, blue not-underlined, blue underlined in degraded and un-degraded versions) with the embedded target word *Nugat* (English: *nougat*). Before each target the gray line indicated the invisible boundary, which triggered the display change from degraded to un-degraded presentations in case a saccade crossed the boundary.

(b) Means and standard errors (vertical bars) of skipping probabilities, first fixation durations, gaze durations and total viewing times. Blue dots indicated blue words and black dots black words. *UL* indicates underlined presentation and *N* indicates normal presentation.

*Table 1*. Fixed effects of LMM analyses for skipping probability, first fixation duration, gaze duration and total viewing time (all timing measures were log transformed).

|  | Fixed effects | SE |  |
|---|---|---|---|
| *Skipping probability* |  |  | Z-values |
| Degradation (Deg) | 0.08 | 0.12 | 0.67 |
| Color (Col) | -0.20 | 0.14 | 1.43 |
| Underlined (Undl) | -0.04 | 0.12 | 0.32 |
| Deg X Col | 0.22 | 0.17 | 1.25 |
| Deg X Undl | 0.18 | 0.17 | 1.09 |
| Col X Undl | 0.05 | 0.18 | 0.30 |
| Deg X Col X Undl | -0.28 | 0.24 | 1.15 |
| *First fixation duration* |  |  | t-values |
| Deg | **-0.110** | **0.015** | **6.73** |
| Col | 0.007 | 0.012 | 0.59 |
| Undl | 0.021 | 0.013 | 1.59 |
| Deg X Col | -0.013 | 0.016 | 0.79 |
| Deg X Undl | -0.015 | 0.017 | 0.91 |
| Col X Undl | -0.012 | 0.017 | 0.69 |
| Deg X Col X Undl | 0.021 | 0.023 | 0.89 |
| *Gaze duration* |  |  |  |
| Deg | **-0.072** | **0.012** | **6.12** |
| Col | 0.004 | 0.011 | 0.33 |
| Undl | -0.001 | 0.010 | 0.10 |
| Deg X Col | -0.009 | 0.014 | 0.65 |
| Deg X Undl | 0.004 | 0.014 | 0.24 |
| Col X Undl | 0.007 | 0.014 | 0.50 |
| Deg X Col X Undl | 0.011 | 0.020 | 0.54 |
| *Total viewing time* | | | |
| Deg | **-0.080** | **0.016** | **4.95** |
| Col | 0.030 | 0.017 | 1.77 |
| Undl | **0.046** | **0.018** | **2.59** |
| Deg X Col | -0.011 | 0.021 | 0.53 |
| Deg X Undl | -0.028 | 0.021 | 1.34 |
| Col X Undl | **-0.044** | **0.021** | **2.09** |
| Deg X Col X Undl | 0.040 | 0.030 | 1.33 |

*Note*. Reliable effects are highlighted in bold numerals

As expected skipping probability, presented in Figure 1b, was not reliably affected by color, underlining or degradation (see Table 1). The present study reports low skipping probabilities between 8 and 12%, when compared to the Fitzsimmons study with skipping probabilities up to 27%. This indicates that fixation rates of the target words are comparable indicating a floor effect for cognitive influences on word skipping.

In contrast, eye movement measures based on fixation durations during first pass reading indicated a strong preview benefit but no effect of color or highlighting. This was shown by the reliably lower first fixation durations and gaze durations for un-degraded parafoveal presentation in contrast to degraded previews (see Figure 1b). No reliable effects and interactions of color or underlining were found (see Table 1). This finding indicates that bottom-up perceptual processing preceding word recognition (i.e., in parafoveal vision) was only influenced by degraded parafoveal previews but not reliably by word color or underlining.

The total viewing times, including re-fixation times after the first encounter (i.e., re-reading), showed, in addition to a reliable degradation effect, a reliable interaction of word color and underlining. Figure 1b clearly shows the origin of this interaction: un-highlighted black-presented words had reduced total viewing times in contrast to all other conditions including blue underlined, blue not-underlined and black underlined words (confirmed by post-hoc analysis: underlining effect for black targets; estimate = 0.046; SE = 0.020; $t$ = 2.29; no underlining effect for blue targets; estimate = 0.002; SE = 0.016; $t$ = 0.12). This indicates that highlighting either by color or underlining increases the re-reading times reflecting the allocation of additional attentional resources to highlighted words after first pass reading. In addition, the reduced skipping probability of blue target words, described by Fitzsimmons and colleagues (2013), might also reflect a highlighting effect for sentences in which target word skipping can be realized to a higher extent.

In sum, the present study demonstrated that reading was not hampered by blue text presentation. Thus, the current findings do not indicate a bottom-up perceptual disadvantage of blue underlined hypertext in foveal and parafoveal processing. In contrast, the increased total viewing time for highlighted stimuli indicates an additional allocation of attentional resources triggered by top-down processes. These processes might reflect the learned association of hypertext to informative snips of texts in the Internet. For now I can only offer congratulations for those who were able to produce such a successful educated guess. Using blue underlined stimuli effectively highlights hypertext without hindering (parafoveal and foveal) perceptual processes during reading. In conclusion, the blue underlined hypertext implementation allows effective reading and, therefore, can be safely recommended to web designers for future use.

Acknowledgements

I want to thank Arturo Hernandez for helpful discussions and comments on an earlier version of the manuscript. I also want to thank Susanne Eisenhauer, Kirsten Hilger and Edvard Heikel for comments on an earlier version and Agnes Altmanninger, Pia Schweitzer and Eva Daspelgruber for helping with the data acquisition. Finally, I want to thank Florian Hutzler for letting me use his laboratory and Stefan Hawelka since part of the current software used in the project was scripted in collaboration for previous studies. This research was supported by the Goethe University Frankfurt "Nachwuchswissenschafter im Fokus – Förderlinie A".